\documentclass[a4paper,11pt]{article}
\usepackage{pos}
\usepackage{units}

\usepackage{xspace}

\newcommand{\Xmax}{$X_{\rm max}$\xspace} 
\newcommand{\Xmaxmath}{X_{\rm max}}

\newlength{\bibitemsep}
\newlength{\bibparskip}
\let\oldthebibliography\thebibliography
\renewcommand\thebibliography[1]{%
  \oldthebibliography{#1}%
  \setlength{\parskip}{\bibitemsep}%
  \setlength{\itemsep}{\bibparskip}%
}

\title{Prospects for measuring the longitudinal particle distribution of cosmic-ray air showers with SKA}
 \ShortTitle{Prospects for measuring cosmic rays with SKA}

\author*[a,b]{A.~Corstanje}
\author[a,b]{S.~Buitink}
\author[j]{J.~Bhavani}
\author[a]{M.~Desmet}
\author[b, c, d]{H.~Falcke}
\author[e]{B.M.~Hare}

\author[b,d,a]{J.R.~H\"orandel}
\author[f,a]{T.~Huege}
\author[f]{N.~Karasthatis}
\author[a]{G.~K.~Krampah}
\author[a]{P.~Mitra}
\author[a]{K.~Mulrey}

\author[g,h]{A.~Nelles}
\author[b]{K.~Nivedita}
\author[a]{H.~Pandya}
\author[a]{J.~P.~Rachen}

\author[i]{O.~Scholten}
\author[j]{S.~Thoudam}

\author[k]{G.~Trinh}

\author[c]{S.~ter Veen}

\affiliation[a]{Vrije Universiteit Brussel, Astrophysical Institute, Pleinlaan 2, 1050 Brussels, Belgium}

\affiliation[b]{Department of Astrophysics/IMAPP, Radboud University Nijmegen\\
 P.O. Box 9010, 6500 GL Nijmegen, The Netherlands}

\affiliation[c]{Netherlands Institute for Radio Astronomy (ASTRON)\\
 Postbus 2, 7990 AA Dwingeloo, The Netherlands}

\affiliation[d]{Nikhef, Science Park Amsterdam, 1098 XG Amsterdam, The Netherlands}
\affiliation[e]{University of Groningen, Kapteyn Astronomical Institute, Groningen, 9747 AD, Netherlands}
\affiliation[f]{Institut f\"{u}r Astroteilchenphysik, Karlsruhe Institute of Technology (KIT) \\
 P.O. Box 3640, 76021, Karlsruhe, Germany}
\affiliation[g]{DESY, Platanenallee 6, 15738 Zeuthen, Germany}
\affiliation[h]{ECAP, Friedrich-Alexander-University Erlangen-N\"{u}rnberg, 91058 Erlangen, Germany}
\affiliation[i]{Interuniversity Institute for High-Energy, Vrije Universiteit Brussel \\
 Pleinlaan 2, 1050 Brussels, Belgium}
\affiliation[j]{Department of Physics, Khalifa University, P.O.~Box~127788, Abu Dhabi, United Arab Emirates}
\affiliation[k]{Department of Physics, School of Education, Can Tho University Campus II \\
 3/2 Street, Ninh Kieu District, Can Tho City, Vietnam}

\emailAdd{a.corstanje@astro.ru.nl}

\abstract{We explore the possibilities of measuring the longitudinal profile of individual air showers beyond \Xmax when using very dense radio arrays such as SKA.
The low-frequency part of the Square Kilometre Array, to be built in Australia, features an enormous antenna density of about 50,000 antennas in the inner core region of radius 500 m, with a frequency band from 50 to 350 MHz. From CoREAS simulations, a SKA-Low antenna model plus noise contributions, and adapted LOFAR analysis scripts, we obtain a resolution in the shower maximum \Xmax and energy that is considerably better than at LOFAR.
Already from this setup, we show that at least one additional parameter of the longitudinal profile can be measured.
This would improve mass composition analysis by measuring an additional composition-dependent quantity. Moreover, it would offer an opportunity to discriminate between the different predictions of hadronic interaction models, hence contributing to hadronic physics at energy levels beyond man-made accelerators.
}
\FullConference{%
  9th International Workshop on Acoustic and Radio EeV Neutrino Detection Activities - ARENA2022\\
  7-10 June 2022\\
  Santiago de Compostela, Spain}

\begin{document}
\maketitle

\section{Introduction}
The low-frequency part of the Square Kilometre Array \cite{Tan:2015} to be built in Australia will feature an unprecedented number and density of antennas. Whereas LOFAR \cite{vanHaarlem:2013} has about 300 usable antennas in a diameter of $\unit[320]{m}$, SKA will have about $50,000$ in a radius of $\unit[500]{m}$, and extends the frequency band to $\unit[350]{MHz}$.
Hence, it is expected to open up new possibilities in measuring air showers into fine detail. 

LOFAR has been successful in measuring \Xmax to an accuracy suitable for cosmic-ray mass composition studies \cite{Corstanje:2021,Buitink:2016}.
The longitudinal evolution of the number of particles in an air shower, reaching a maximum at \Xmax, has two extra parameters containing information about mass composition; moreover, their expected value and distribution varies with hadronic interaction model \cite{Buitink:2022}. Hence, being able to measure these would offer an opportunity to improve mass composition measurements, and to distinguish between hadronic interaction models at high energies.

Here we report a first analysis to reconstruct longitudinal profile parameters.

\section{Methods}
\begin{figure}
	\centering
	\includegraphics[width=0.40\textwidth]{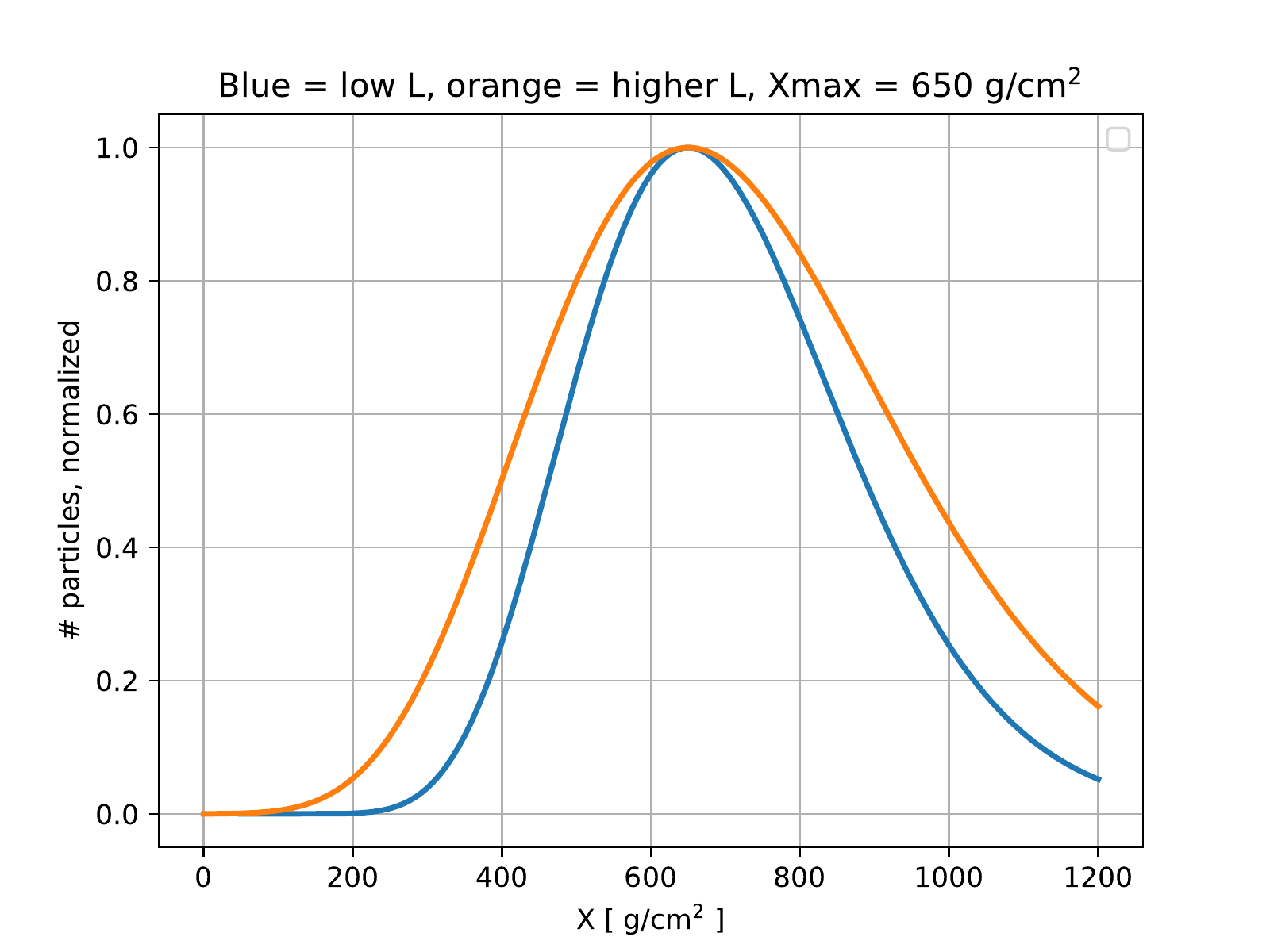}
	\includegraphics[width=0.40\textwidth]{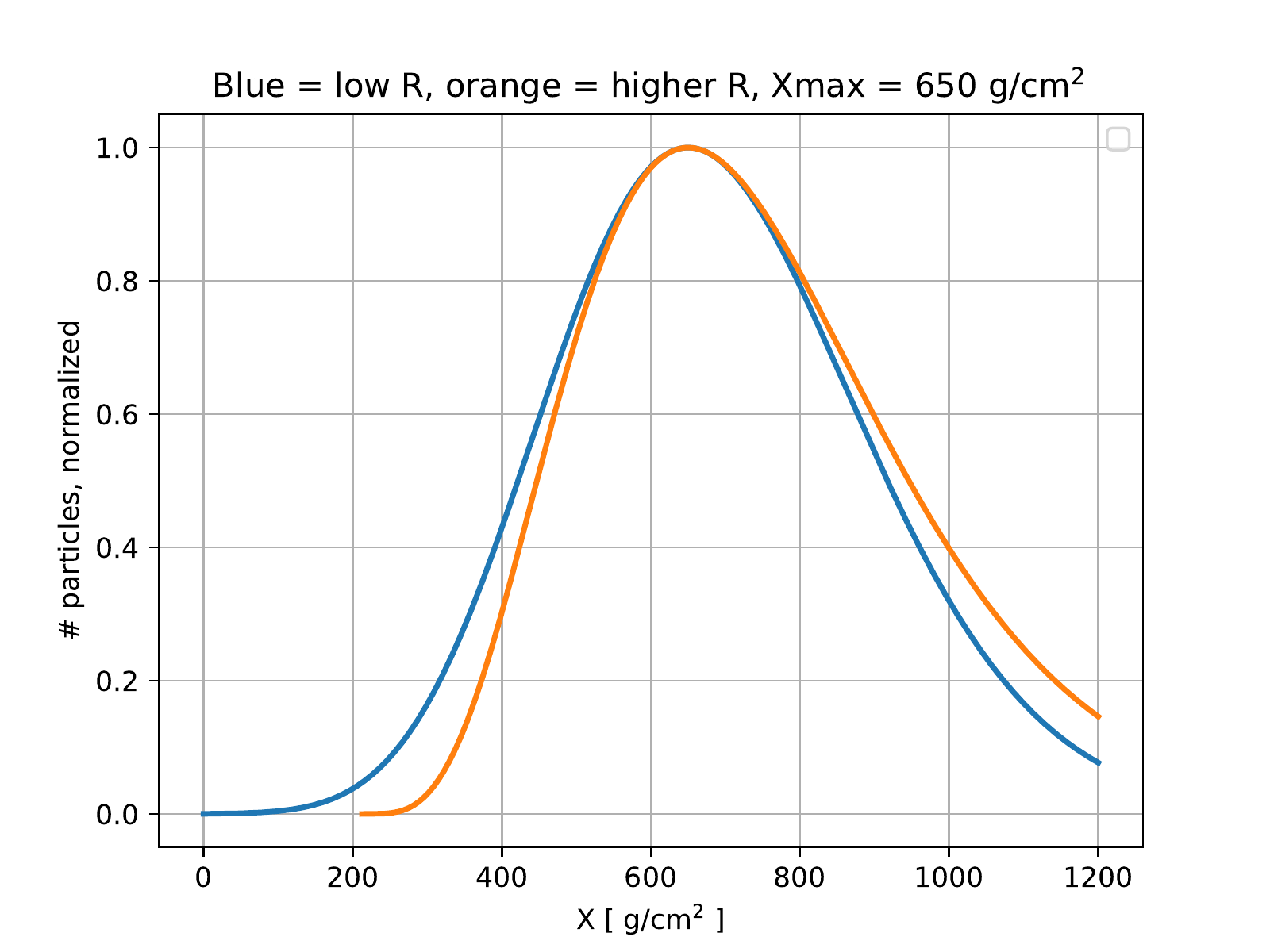}
	\caption{An example of a longitudinal particle distribution at $\Xmaxmath=\unit[650]{g/cm^2}$, showing the effect of varying $L$ (left) and $R$ (right). The variations of $L$ and $R$ have been exaggerated compared to typical natural values.}
\label{fig:longdist}
\end{figure}

The longitudinal particle distribution of air showers, of which an example is shown in Fig.~\ref{fig:longdist}, can be parametrized by three parameters \Xmax, $L$, and $R$ \cite{Andringa:2011}.
The distribution function is
\begin{equation}\label{eq:LRdist}
N(X) = \exp \left(-\frac{X - \Xmaxmath}{RL}\right)\,\left(1 - \frac{R}{L}\left(X - \Xmaxmath\right)\right)^{\frac{1}{R^2}},
\end{equation}
with $N$ denoting the number of particles, and $X$ the traversed depth in the atmosphere in $\unit[]{g/cm^2}$.
In this distribution, \Xmax is the maximum or {\it mode}, $L$ is related to the variance (second moment), and $R$ to the skewness (third moment). 
The effect of varying $L$ and $R$ is demonstrated in Fig.~\ref{fig:longdist}.

In radio measurements at LOFAR, we have focused on reconstructing \Xmax, as this is the parameter to which the radio footprint is most sensitive.
To investigate the possibilities of reconstructing $L$ and $R$ with SKA, we have simulated an ensemble of $110$ proton showers at an energy of $\unit[10^{17}]{eV}$, coming from east at a zenith angle of 30~degrees. Simulations were done with Corsika and CoREAS 7.7410 \cite{Corsika:1998, CoREAS:2013} at a thinning level of $10^{-6}$.
Using pre-selection of showers with CONEX \cite{Bergmann:2007}, these showers were selected to have an \Xmax (from Corsika) of $\unit[645]{g/cm^2}$, all within a range of $\pm \unit[0.5]{g/cm^2}$. For this, simulating about 50000 Conex showers and 500 Corsika showers is sufficient for producing the $110$ showers with Corsika plus CoREAS.
Having effectively taken out the \Xmax-variations, we can focus on the effect of variations of $L$ and $R$ on the radio footprint.

The present analysis focuses on two parts: comparing radio footprints directly as a function of $L$ and $R$, and doing a more complete reconstruction of \Xmax, $L$, and $R$ using simulated measurements with realistic sky noise.

\subsection{Interpolation based on Fourier series}\label{sect:interpolation}
Our simulations have antennas put on a radial grid or `star shape', with radial density higher in the first $\unit[100]{m}$ from the core, then declining outward.
For a reconstruction from a dense layout of antennas, an interpolation is needed.
We have developed an interpolation method that improves over the radial basis function method \cite{Scipy:2020} used earlier, which does not perform as well on footprints of higher-frequency signals.
It makes use of the natural radial structure of the footprint, and of the (mainly) $\cos(\phi)$-dependence of the pulse energy from geomagnetic and charge-excess origin combined.

On a radial grid, the energy along the angular direction (at constant radius) can be expressed as a Fourier series, by taking an FFT.
In our case, the FFT is done on 8 points (simulated antennas) along a circle at each radius, and for convenience we express the result in cosine and sine amplitudes, here denoted by $c_k(r)$ and $s_k(r)$ for mode $k$ at radius $r$.
This is a complete representation of the values along a circle, provided it is sampled densely enough (the Nyquist criterion on spatial frequencies).

The interpolated footprint energy $f(r, \phi)$ at position $\phi$ along a circle of radius $r$ is given by
\begin{equation}\label{eq:interpolation_def}
f(r, \phi) = \sum_{k=0}^{n/2} c_k(r)\cos(k\,\phi) + s_k(r)\sin(k\, \phi).
\end{equation}
For interpolation in the radial direction, a standard one-dimensional interpolation is used on the angular Fourier components $c_k(r)$ and $s_k(r)$ separately. A cubic spline interpolation was found useful here.
After this, Eq.~\ref{eq:interpolation_def} can be used at any position within the radial boundaries of the simulated antennas.
Using this method, simulations having on the order of $200$ antennas would suffice for producing pulse energy footprints to an accuracy as desired  for application to SKA measurements. 
A dedicated publication on this method and its precision is forthcoming.

\subsection{Direct comparison of equal-\Xmax footprints as a function of $L$ and $R$}
A direct comparison between footprints is made by determining the mean squared error (MSE) when taking their difference on each simulated antenna position, fitting a constant scale factor.
This amounts to
\begin{equation}\label{eq:mse}
\mathrm{MSE} = \sum_{\mathrm{ant}} \left(f_i(\vec{x}) - A f_j(\vec{x})\right)^2,
\end{equation}
with the best-fit energy scale factor $A$, and $f_i$ the pulse energies in shower $i$.
The MSE is the `noiseless' equivalent of the $\chi^2$ used when fitting simulations to realistic data with noise.

A second way of comparing footprints is to consider the (lowest-order) Fourier amplitudes as a function of distance to shower core.
Taking one shower as reference, e.g.~around the median value of $L$ in the ensemble, we take the difference between showers of the angular-constant and $\cos(\phi)$ modes, respectively.

\subsection{Monte Carlo setup for simulated air shower reconstruction at SKA}
To mimic the essential steps of the reconstruction process at SKA, we make use of the simulated ensemble of CoREAS showers to create mock data. The signals at each simulated antenna are passed through the SKALA2 antenna model \cite{Acedo:2015}, giving voltages as they would be seen at the AD converters.
Realistic levels of (frequency-dependent) sky noise are added to the simulated pulse signal at each antenna on the radial grid.
The power spectrum of sky noise is taken from LFMap \cite{Polisensky:2007} as was used for calibration of the LOFAR antennas \cite{Mulrey:2019}.
We scale the simulated pulse amplitudes, to mimic an energy of $2 \times \unit[10^{17}]{eV}$, to increase the signal-to-noise ratio in a restricted band of 50 to $\unit[100]{MHz}$.

The pulse energy is measured in a time window around the pulse maximum, with a width of $\unit[12]{ns}$ for the full $50$ to $\unit[350]{MHz}$ bandwidth, or $\unit[60]{ns}$ when restricting to $50$ to $\unit[100]{MHz}$. A (non-tight) trigger criterion of $5\,\sigma$ above the noise is used to obtain a trigger probability versus pulse energy.
By taking 300 noise realisations, we obtain an uncertainty on measured pulse energy at each antenna position.
Now, we use the interpolation method from Sect.~\ref{sect:interpolation} to obtain pulse energy and its uncertainty at each SKA antenna position in the proposed layout.

The reconstruction process proceeds as is done at LOFAR \cite{Corstanje:2021,Buitink:2014}, with minimally adapted code.
For the near-constant-\Xmax ensemble we try reconstruction of $L$ and of a combination of $L$ and $R$ described below.

\section{Results of direct comparison}
\begin{figure}
	\includegraphics[width=0.50\textwidth]{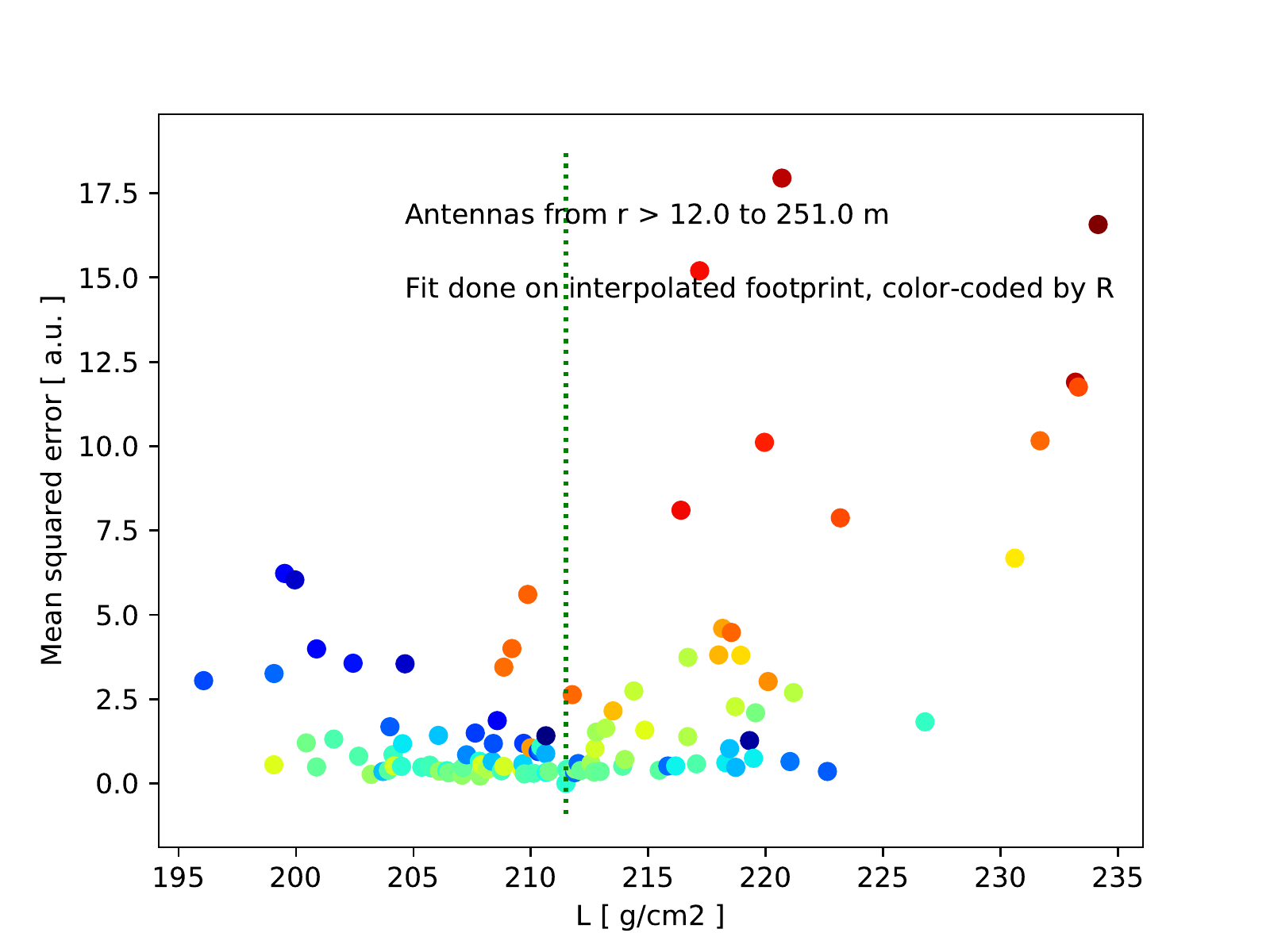}
	\includegraphics[width=0.50\textwidth]{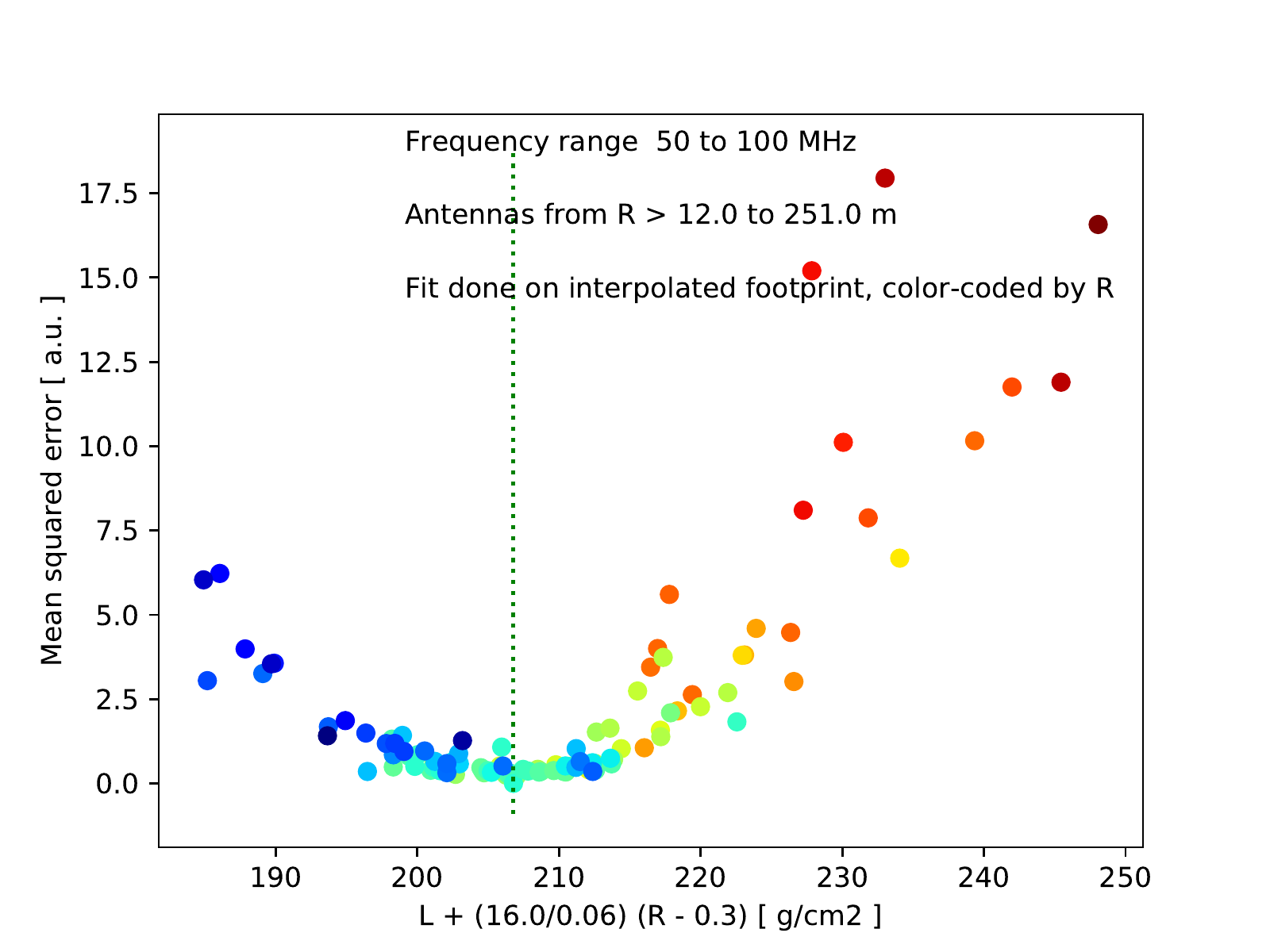}
	\caption{Left: mean squared difference between the ensemble footprints and a test shower at $L=\unit[212]{g/cm^2}$, color-coded by $R$. 
	Right: the same, but expressed as a function of a linear combination of $L$ and $R$.}
\label{fig:comparison}
\end{figure}

We have taken a shower around the median value of $L$ in our ensemble, and computed the MSE when fitting each of the other showers to it.
We have restricted the frequency range to $50$ to $\unit[100]{MHz}$ as the effects shown below appear most clearly in this lower frequency band.
The result is shown in the left panel of Fig.~\ref{fig:comparison}, as a function of $L$ and color-coded by $R$. 
It is seen that no clear minimum arises, but there appears to be a non-random variation with $R$.

Therefore, we consider expressing the MSE as a function of a combination of $L$ and $R$, obtaining the right panel of Fig.~\ref{fig:comparison}.
The simplest combination is a linear function of $L$ and $R$, with a constant coefficient to put them on the same dimension of $\unit{g/cm^2}$. We introduce a sensitive parameter $S$ as
\begin{equation}\label{eq:LRcombi}
S(L, R) = L + \frac{\unit[16]{g/cm^2}}{0.06} \left(R - 0.3\right),
\end{equation}
where the numbers $0.06$ and $0.3$ are approximately the half-range and the median value of $R$, respectively.
The coefficient value of $\unit[16]{g/cm^2}$ depends on the frequency range, and possibly on other shower parameters such as \Xmax and zenith angle which were kept constant here.
This shows a clear minimum around the correct position, and (to lowest order) a parabolic shape around it, as familiar from \Xmax reconstructions. 
We conclude that when comparing footprints in this way, they are sensitive to the parameter $S$ rather than to $L$ and $R$ individually.
A way to interpret this is to consider the leading and trailing curves in Fig.~\ref{fig:longdist}, noting that the footprint must be sensitive to the region around \Xmax, hence to a combination of the curve before and after \Xmax. This combination may well be different from the definitions of $L$ and $R$ in Eq.~\ref{eq:LRdist}, depending e.g.~on the shower geometry.

\subsection{Angular Fourier modes as function of distance to shower core}
\begin{figure}
	\includegraphics[width=0.50\textwidth]{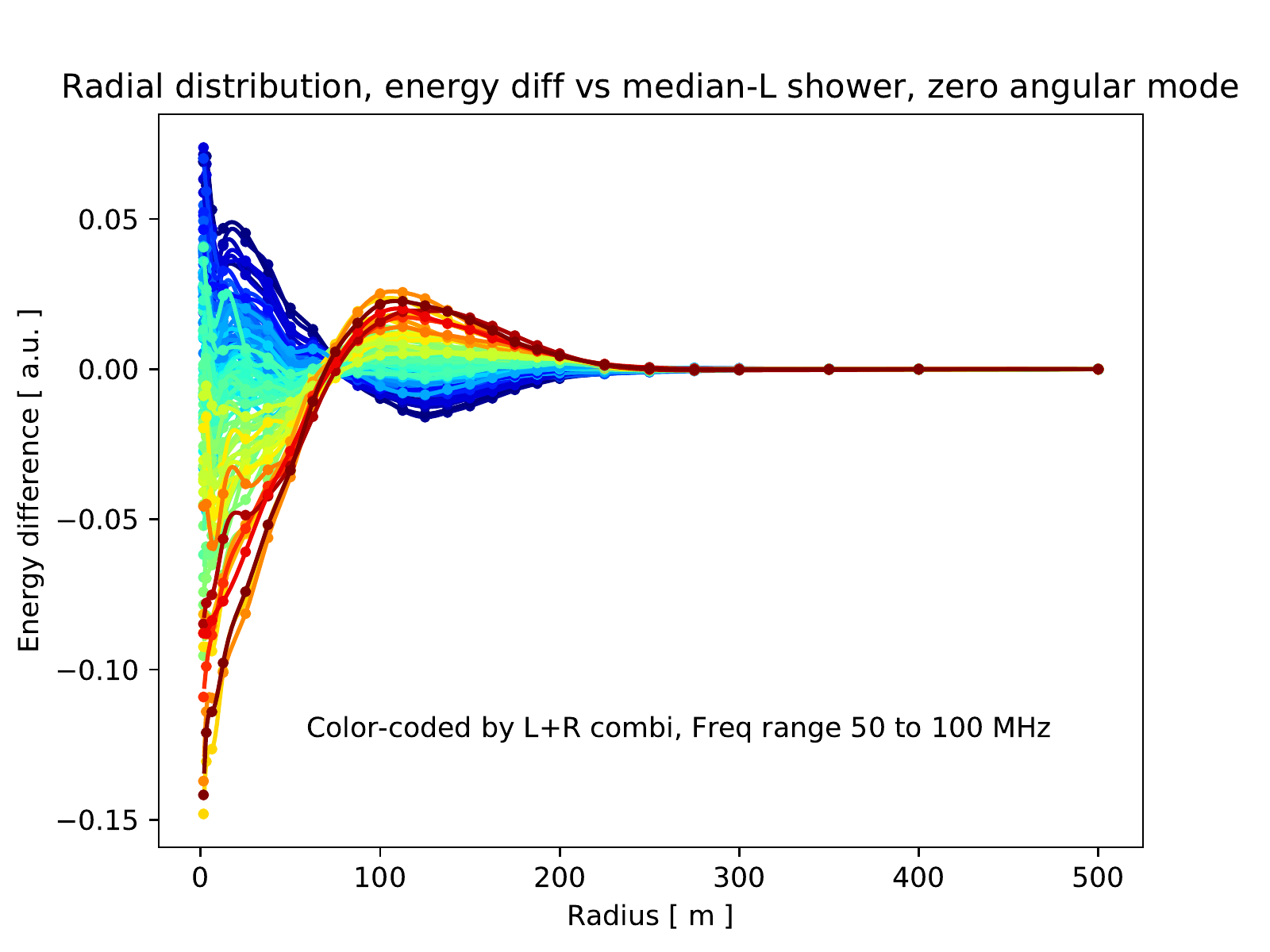}
	\includegraphics[width=0.50\textwidth]{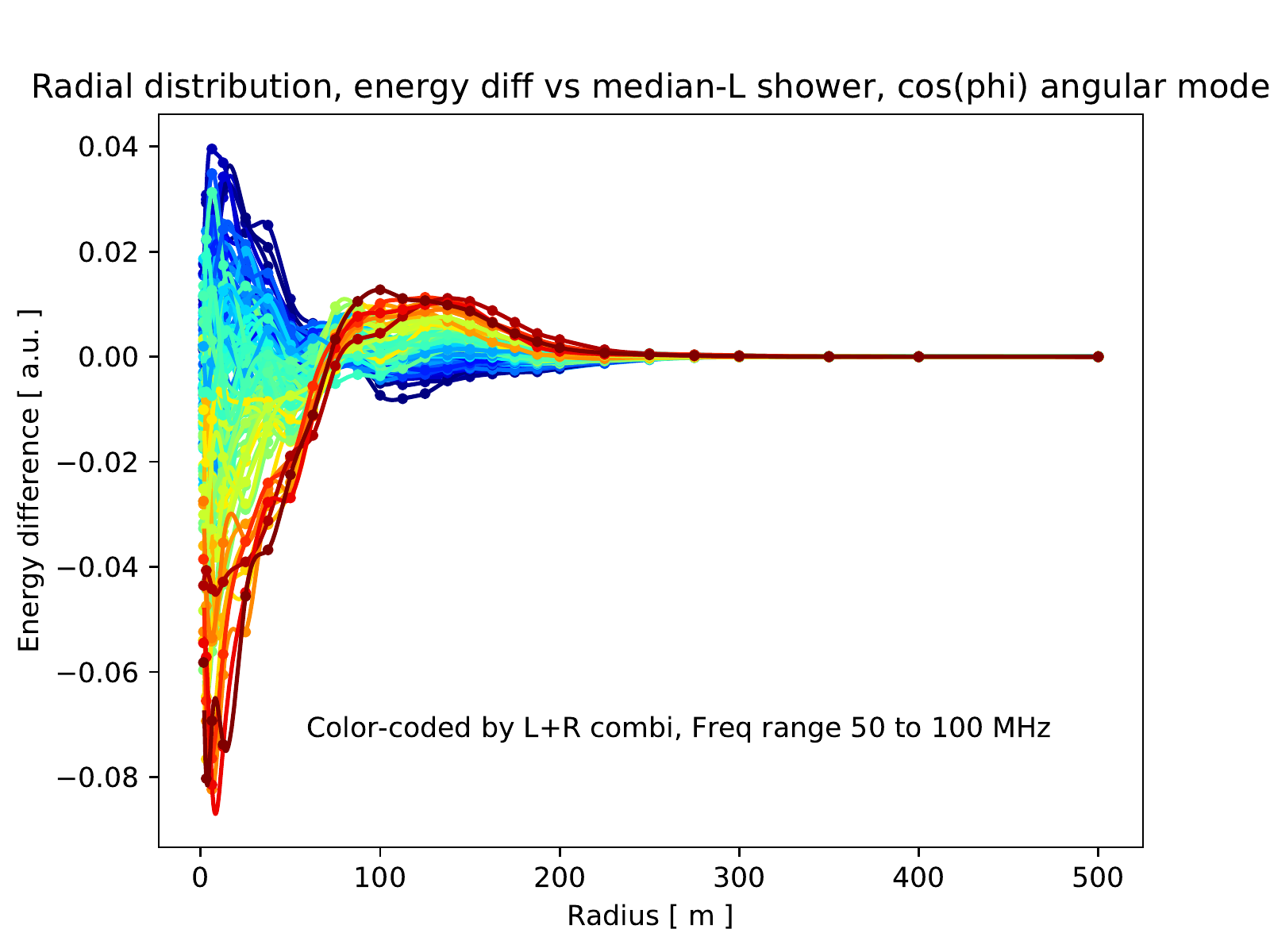}
	\caption{Left: constant angular energy as a function of distance $r$ to shower core, color-coded by the parameter $S$ or `$L/R$ combination' found in Eq.~\ref{eq:LRcombi}. Right: the same for the $\cos(\phi)$ mode.}
\label{fig:fouriermodes}
\end{figure}

In Fig.~\ref{fig:fouriermodes} we look at the two leading Fourier modes, being the angular-constant and $\cos(\phi)$-mode. 
The median-$L$ shower is taken as baseline, and the difference of all other showers with respect to this shower is plotted, color-coded by the parameter $S$ described above.
It is seen from the left panel that at low values of $S$, the footprint has relatively more energy in the inner part, at a core distance below about $\unit[80]{m}$. The right panel shows that additionally, the angular dependence is stronger for these showers as well in the inner part, both in the absolute sense as plotted here, and relative to the (dominant) angular-constant mode.

Thus, seeing a clear difference in footprint shape, we conclude that the footprints contain measurable information on at least the parameter $S$. If this is not fully degenerate with the changes in footprint shape from a varying \Xmax, this must be measurable in practice with a very dense array such as SKA.

\section{Result of simulated reconstruction}
We show the result of a typical simulated reconstruction of \Xmax in the full frequency range, and of the parameter $S$.
The former is done on an ensemble spanning a range of $\pm \unit[25]{g/cm^2}$ in \Xmax; the latter is done on the ensemble at constant \Xmax $\pm \unit[0.5]{g/cm^2}$.

\subsection{Reconstruction of \Xmax in the 50 to 350 MHz frequency band}
\begin{figure}
	\centering
	\includegraphics[trim=2cm 0 13.9cm 0, clip, width=0.6\textwidth]{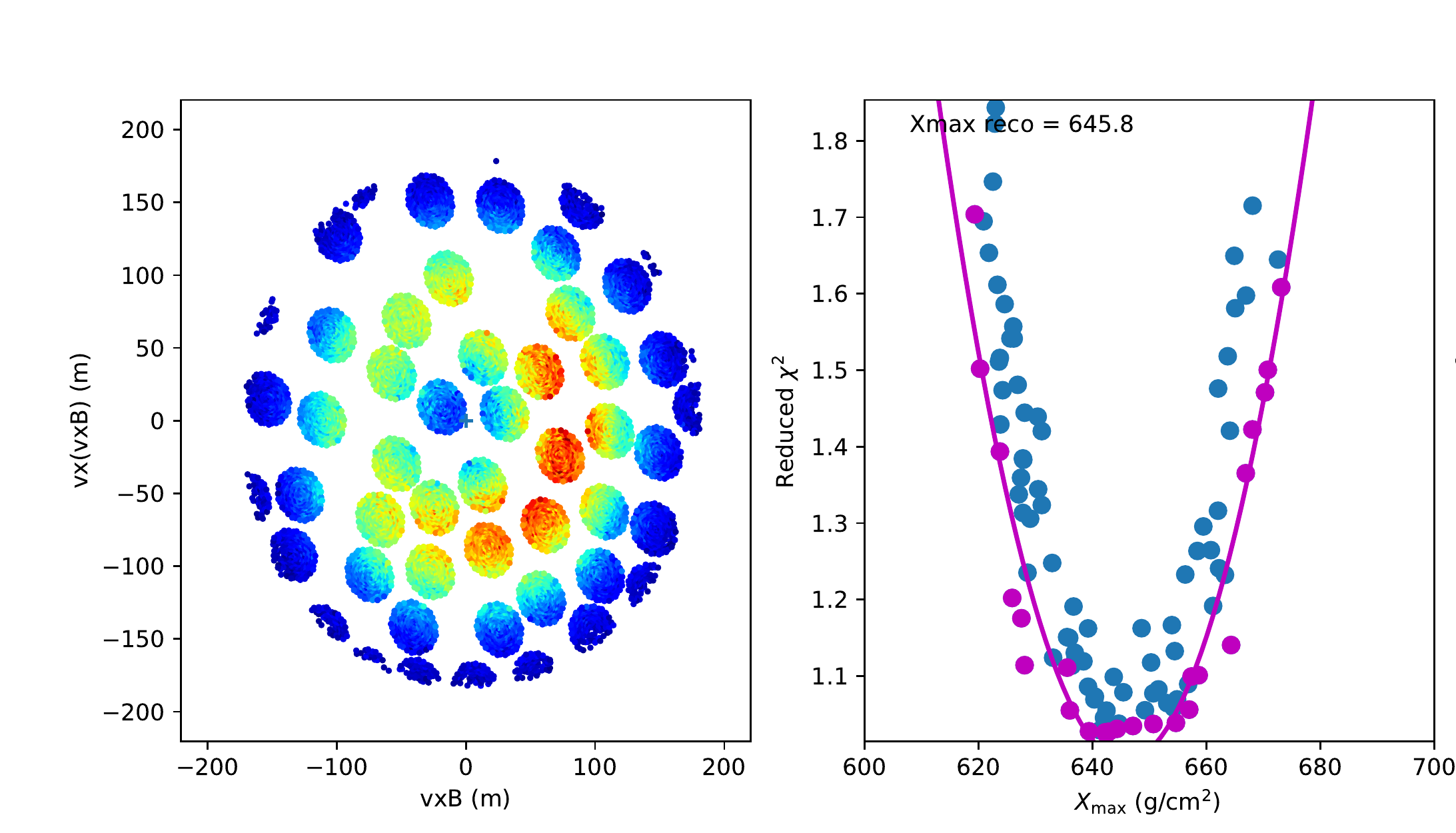}
	\caption{Left: pulse energy in antenna positions projected onto the shower plane, from simulated measurement with noise. Right: reduced $\chi^2$ of fitting ensemble showers to the measured shower. The parabola is fitted to the points shown in magenta forming a lower envelope to the data points.}
\label{fig:xmax_reco_example}
\end{figure}

The \Xmax-reconstruction is shown in Fig.~\ref{fig:xmax_reco_example}. Variations in $L$ and $R$ have not been used here, as in the LOFAR \Xmax-reconstruction.
It gives a fitted value of \Xmax that is within $\unit[1]{g/cm^2}$ of the true value.
A typical accuracy for \Xmax reconstructed in this way is 6 to $\unit[8]{g/cm^2}$.

\subsection{Reconstruction of a sensitive parameter $S$ in the 50 to 100 MHz band}
\begin{figure}
	\centering
	\includegraphics[trim=2cm 0 14.12cm 0, clip, width=0.6\textwidth]{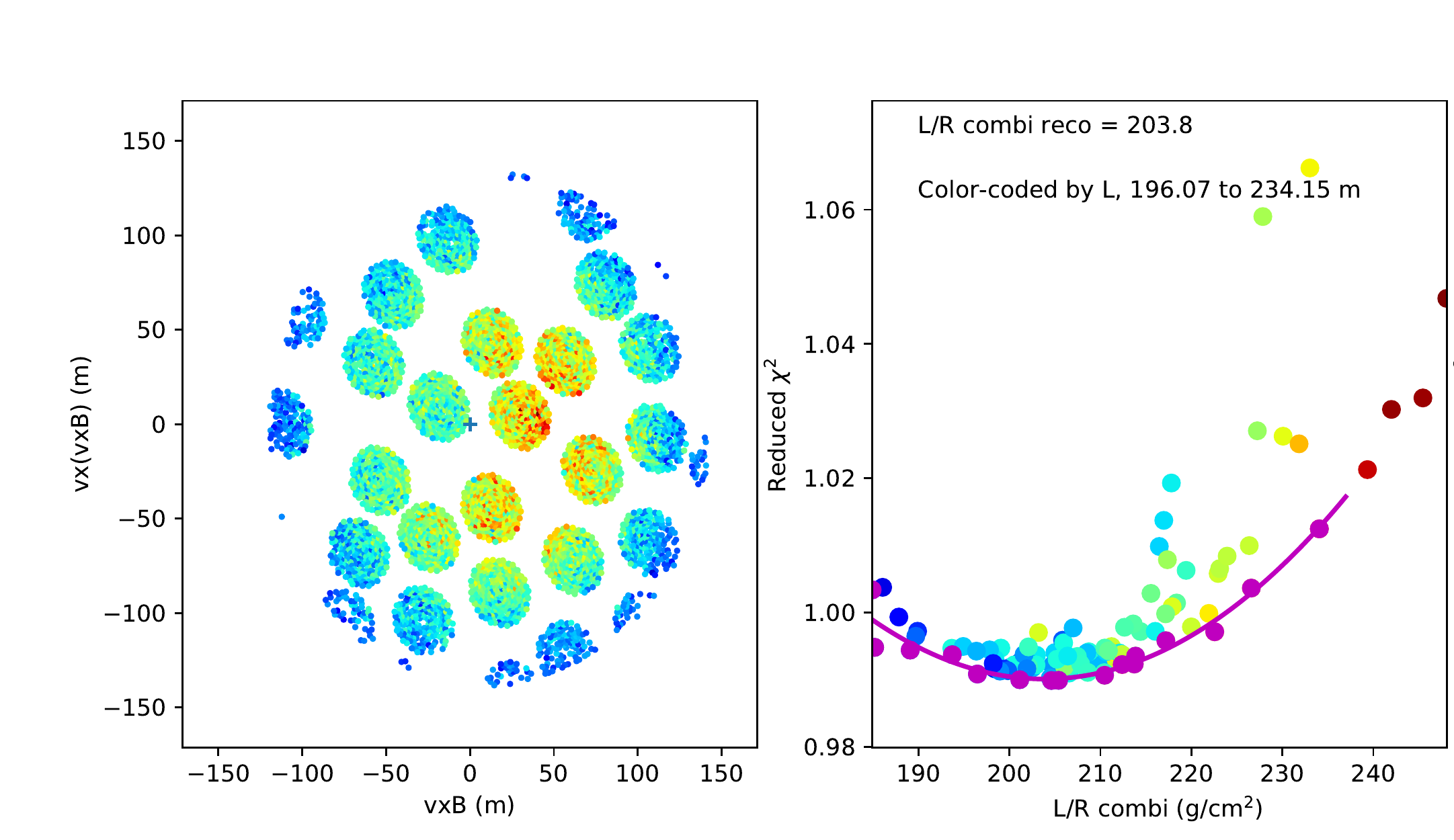}
	\caption{Same as in Fig.~\ref{fig:xmax_reco_example}; color-coding in the right panel corresponds to the value of $L$}
\label{fig:LR_reco_example}
\end{figure}

In Fig.~\ref{fig:LR_reco_example} we show the result of reconstructing the parameter $S$ from equation \ref{eq:LRcombi}.
Note that the reconstruction was done with the fixed-\Xmax ensemble.
Again, the $\chi^2$ versus $S$ shows a clear minimum, well fitted by a parabola. The fitted value is close to the real value of $\unit[204.6]{g/cm^2}$.

\begin{figure}
	\includegraphics[width=0.33\textwidth]{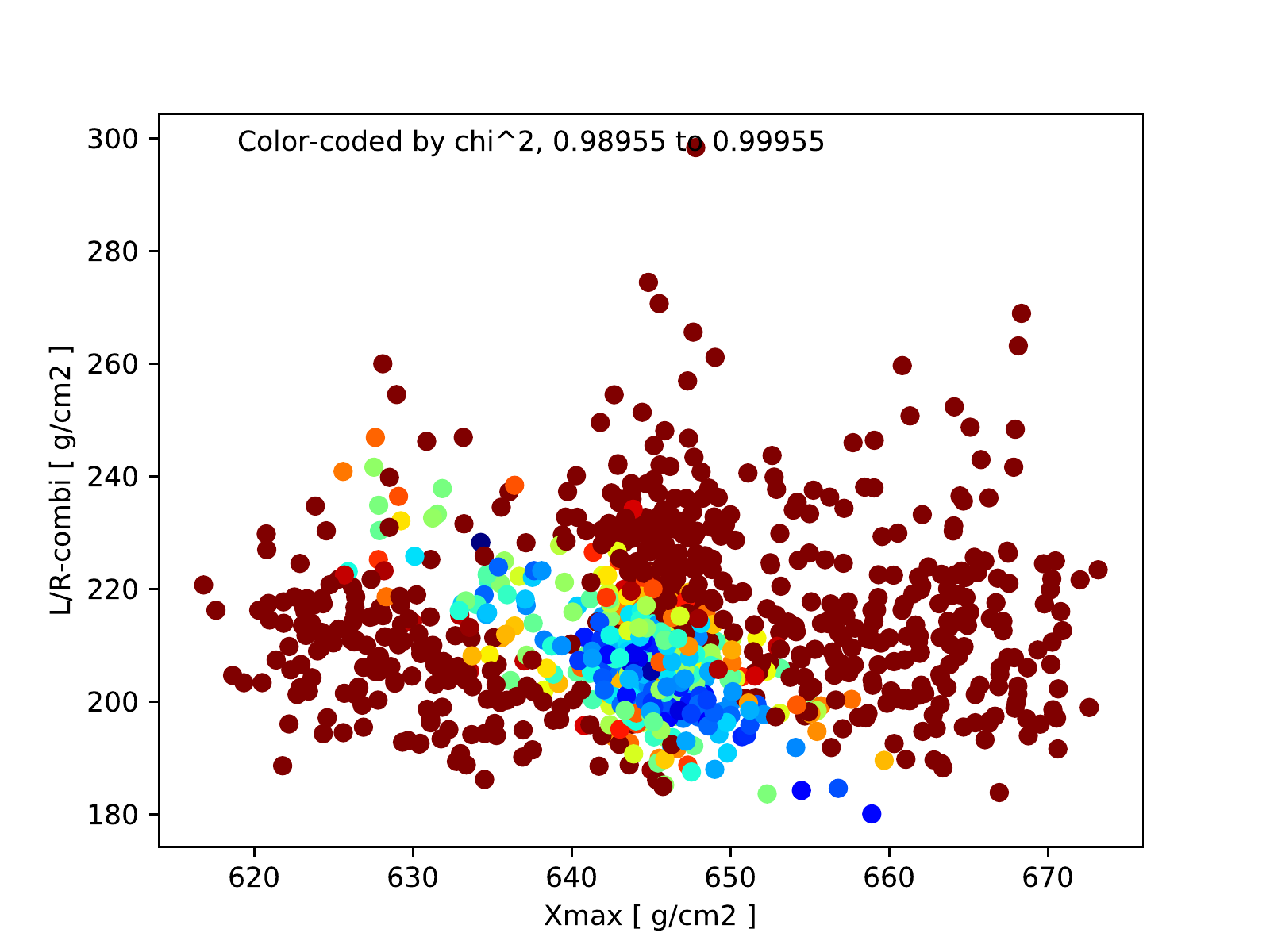}
	\includegraphics[width=0.33\textwidth]{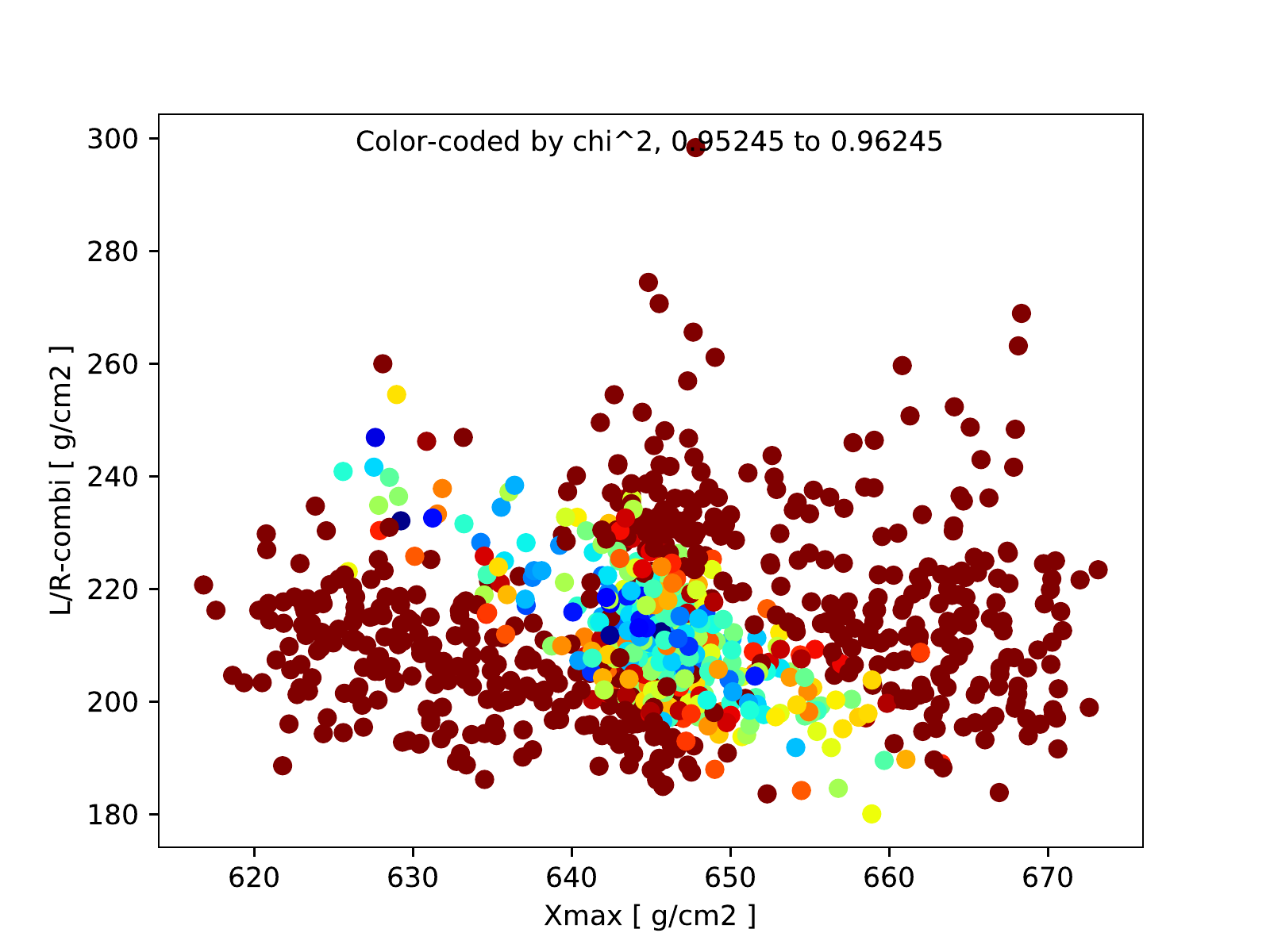}
	\includegraphics[width=0.33\textwidth]{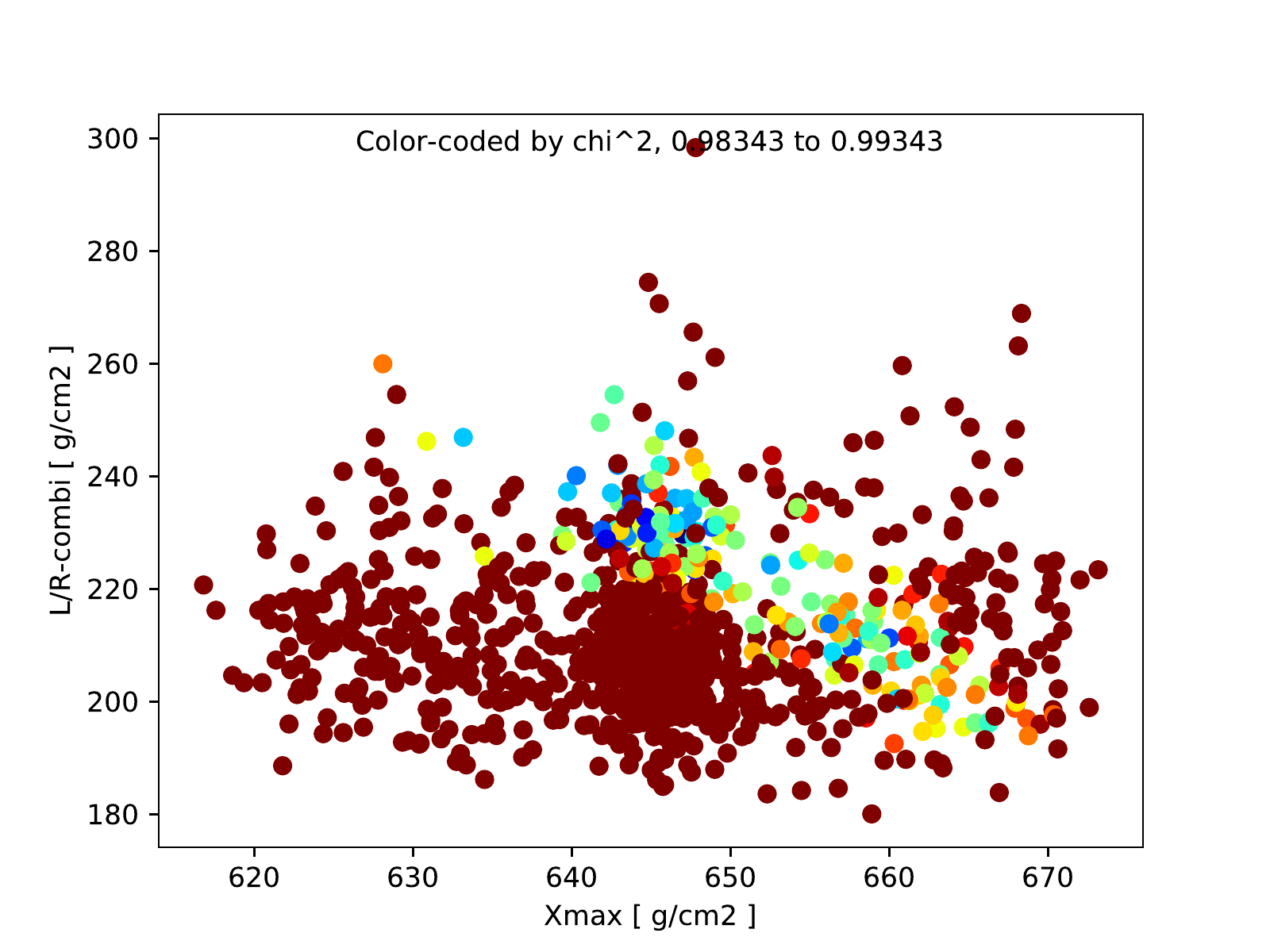}
	\caption{Fit quality (color-coded) as a function of \Xmax and the parameter $S$ from Eq.~\ref{eq:LRcombi}, for a low, medium, and high value of $S$, respectively. A region of near-optimal fits is seen along a slanted line in these plots.}
\label{fig:LR_Xmax_both}
\end{figure}

In a realistic measurement, we have to infer \Xmax, $L$, and $R$ (or a combination of them) simultaneously.
For a first test of the capabilities to distinguish them, we have simulated additional showers in a wider \Xmax range, and plotted their value of $S$ versus \Xmax, color-coded by the reduced $\chi^2$ of their fit. 
Results for three showers are shown in Fig.~\ref{fig:LR_Xmax_both}; the color scale has been capped at a value not far above the minimum, to demonstrate the region where near-optimal fits are found.
We find a `valley' along a slanted line in these plots, where a higher \Xmax and a lower $S$ or vice versa can produce nearly the same fit quality close to the optimum.
Nevertheless, having determined \Xmax from the full frequency band to about $\pm \unit[7]{g/cm^2}$, we can already distinguish low, medium, and high values of $S(L,R)$.
Extending this analysis to the full bandwidth has the potential of significantly improving on this result.

\section{Summary}
We have demonstrated cosmic-ray air shower measurement capabilities of SKA beyond what is possible at current radio observatories.
To this end we have simulated the measurement process by passing simulated radio signals through the SKALA2 antenna model, adding realistic sky noise levels, and measuring pulse energies around their maximum.

In the three-parameter description of the longitudinal distribution of particles by \Xmax, $L$, and $R$, we find that the radio footprint is sensitive to \Xmax (as was already known) and to a linear combination of $L$ and $R$ which we denoted by $S$. Using an ensemble of showers at a fixed \Xmax, and restricting the bandwidth to $50$ to $\unit[100]{MHz}$, we have seen that this combination of $L$ and $R$ can be reliably reconstructed in the same way as was already done for \Xmax.
Given a wider ensemble and \Xmax results from the full frequency range, we can in this limited setup already distinguish between low, medium, and high values of $S$.

The present analysis serves as a first exploration of the capabilities of a very densely instrumented radio telescope like SKA.
Therefore, the results found here are only a lower bound to the precision that can be achieved.
Measuring longitudinal shower parameters beyond \Xmax, for individual showers, is not readily available at cosmic ray observatories in general.
Thus, the ability to do this with SKA would once more demonstrate the value of radio measurements in detailed studies of air showers.

\section*{Acknowledgements}
We acknowledge funding from the European Research Council under the European Union's Horizon 2020 research and innovation programme
(grant agreement n.~640130).
TNGT acknowledges funding from Vietnam National Foundation for Science and Technology Development (NAFOSTED) under grant number 103.01-2019.378. ST acknowledges funding from the Khalifa University Startup grant, project code 8474000237-FSU-2020-13.

\bibliographystyle{jhep}
\bibliography{SKAprospects}

\end{document}